\documentclass[12pt]{article}

\input epsf

\newcommand{\myphantom}[1]{\rule{0pt}{#1}}

\newcommand{\ie}{i.e.\ }

\newcommand{\eq}{eq.~}
\newcommand{\fig}{fig.~}

\newcommand{\diag}{\mathrm{diag}}
\newcommand{\Tr}{\mathrm{Tr}}

\newcommand{\C}{{\sf C\hspace*{-0.9ex}%
    \rule{0.15ex}{1.5ex}\hspace*{0.9ex}}}
\newcommand{\Z}{{\sf Z\hspace*{-0.9ex}%
    \rule{0.15ex}{1.5ex}\hspace*{0.9ex}}}

\newcommand{\sect}[1]{ \section{#1} }

\newcommand{\ve}{\left( \begin{array}{r}}
\newcommand{\ev}{\end{array} \right)}
\newcommand{\ar}{\left( \begin{array}{rr}}
\newcommand{\ra}{\end{array} \right)}
\newcommand{\arr}{\left( \begin{array}{rrrr}}
\newcommand{\arrr}{\left( \begin{array}{rrrrrr}}

\newcommand{\eqr}{\begin{eqnarray}}
\newcommand{\rqe}{\end{eqnarray}}

\newcommand{\half}{\frac{1}{2}}
\newcommand{\third}{\frac{1}{3}}
\newcommand{\quart}{\frac{1}{4}}
\newcommand{\fifth}{\frac{1}{5}}

\def\KK{{\rm I\kern -.2em  K}}
\def\NN{{\rm I\kern -.16em N}}
\def\RR{{\rm I\kern -.2em  R}}

\def\ZZZ{{\small{\rm Z}\kern -.5em Z}}
\def\QQ{{\rm \kern .25em
             \vrule height1.4ex depth-.12ex width.06em\kern-.31em Q}}
\def\CC{{\rm \kern .25em
             \vrule height1.4ex depth-.12ex width.06em\kern-.31em C}}

\title{E$_8$ Quiver Gauge Theory and Mirror Symmetry}
\author{Cecilia Albertsson$^1$\footnote{email: cecilia@physto.se}, 
        Bj\"orn Brinne$^1$\footnote{email: brinne@physto.se}, 
        Ulf Lindstr\"om$^1$\footnote{email: ul@physto.se}
        \\ 
        and 
        \\
        Rikard von Unge$^2$\footnote{email: unge@physics.muni.cz}}
\date{ }

\begin{document}

\maketitle

\begin{center}
\vspace{-1cm}
{\em
$^1$Department of Physics,
Stockholm University\\
Box 6730,
SE-113 85 Stockholm,
Sweden \\
\vspace{0.3cm}
$^2$Institute for Theoretical Physics and
Astrophysics\\ Faculty of Science, Masaryk University\\
Kotl\'{a}\v{r}sk\'{a} 2, CZ-611 37, Brno, Czech Republic\\}

\end{center}

\begin{abstract}
  We show that the Higgs branch of a four-dimensional Yang-Mills
  theory, with gauge and matter content summarised by an $E_8$ quiver
  diagram, is identical to the generalised Coulomb branch of a
  four-dimensional superconformal strongly coupled gauge theory with
  $E_8$ global symmetry. This is the final step in showing that there
  is a Higgs-Coulomb identity of this kind for each of the cases
  $\{0\}$, $A_1$, $A_2$, $D_4$, $E_6$, $E_7$ and $E_8$. This series of
  equivalences suggests the existence of a mirror symmetry between the
  quiver theories and the strongly coupled theories. We also discuss how
  to interpret the parameters of the quiver gauge theory in terms of
  the Hanany-Witten picture.
\end{abstract}

\vspace{-18cm}
\begin{flushright}
USITP-0102\\
February 2001\\
hep-th/0102038\\
\end{flushright}

\thispagestyle{empty}

\newpage

\setcounter{page}{1}

\sect{Introduction}
\label{sec:Intro}

The set of four-dimensional N=2 superconformal $SU(2)$\footnote{Here
  "$SU(2)$" refers to those of the theories that posess a Lagrangian.}
Yang-Mills gauge theories with coupling $\tau = {i \over g^2} +
\theta$ follows Kodaira's classification \cite{Kodaira} of toroidal
singularities ($\tau$ being the torus modulus). (We shall refer to
these theories as Seiberg-Witten (SW) theories). This classification
falls into an ADE pattern, and the type of singularity gives the
global symmetry of the corresponding theory.  There are $A_n$ and
$D_n$ singularities at Im $\tau = \infty$, as well as a $D_4$
singularity that can occur at all values of $\tau$.  Furthermore,
there are six singularities at finite values of $\tau$, of types
$\{0\}$, $A_1$, $A_2$, $E_6$, $E_7$ and $E_8$. Hence there are
precisely seven strongly coupled superconformal theories, with global
symmetries equal to the corresponding singularity type (see also
\cite{Dasgupta1,Seiberg}).  Of these theories the latter three have no
Lagrangian description, which makes them difficult to study.

However, a possible way to sidestep this difficulty appeared when
\cite{E6} found the remarkable fact that the generalised Coulomb
branches of the $\{0\}$, $A_1$, $A_2$, $D_4$ and $E_6$ theories are
identical to the Higgs branches of other, a priori completely
different gauge theories.  These gauge theories, which we will call
\emph{quiver} theories, are four-dimensional N=2 supersymmetric
Yang-Mills theories with gauge and matter content that may be
described by an ADE Dynkin diagram (or, rather, a \emph{quiver}
diagram).  This Dynkin diagram is precisely that of the ADE global
symmetry of the corresponding SW theory.

The authors of \cite{E7} extended the analysis to the $E_7$ case and,
encouraged by these results, proposed the existence of a mirror
symmetry between four-dimensional SW theories and quiver
gauge theories, analogous to the mirror symmetry acting on
three-dimensional gauge theories \cite{IS}. The mirror symmetry would
exchange the generalised Coulomb branch of one theory with the Higgs
branch of the other, and would provide a map between mass parameters
of one theory and Fayet-Iliopoulos parameters of the other.  Since the
Coulomb branch receives quantum corrections but the Higgs branch does
not, one consequence of the mirror symmetry is that quantum effects in
one theory arise classically in the dual theory, and vice versa.

In this paper, we show the remaining moduli space identity, \ie for
the $E_8$ case, thus exhausting the set of SW theories. We also
derive, at least implictly, the map between mass and Fayet-Iliopoulos
parameters for the $E_8$ case.

The quiver gauge theory can be realised as the worldvolume theory of a
D3-brane probing an orbifold singularity, and in Section
\ref{sec:Quiver} we derive the gauge and matter content from such a
setup, showing how to describe it by means of a quiver diagram.  In
Section \ref{sec:E8curve} we construct the curve describing the Higgs
branch, expressed in gauge group invariants and Fayet-Iliopoulos
parameters. In Section \ref{sec:HWpicture} we attempt a deeper
understanding of the quiver theory in terms of the Hanany-Witten
picture \cite{HW}, and finally, Section \ref{sec:Conclusions} contains
a summary and discussion.

\sect{The quiver Higgs branch}
\label{sec:Quiver}

To construct the quiver gauge theory we start from type IIB string
theory in ten flat dimensions (labelled $0, 1, ... , 9$) and make an
orbifold $\C^2/\Gamma$ out of the 6789 directions \cite{JM}.  If
$\Gamma$ is a discrete subgroup of $SU(2)$, then (the non-compact)
$\C^2/\Gamma$, with its single fixed point at the origin, may be
viewed as a local description of a K3 orbifold near one of its several
fixed points.

Then $\Gamma$ may be chosen from one of the following groups: the
cyclic groups $\Z_n$, the dihedral groups ${\cal D}_n$, the trihedral
group ${\cal T}$, the octahedral group ${\cal O}$, and the icosahedral
group ${\cal I}$. Depending on the choice of $\Gamma$, the resulting
quiver theory will be associated with a Dynkin diagram of type $A_n$,
$D_n$, $E_6$, $E_7$ or $E_8$, respectively. As we are interested in
the last case here, we take $\Gamma= {\cal I}$.

We now probe the singularity at the origin of $\C^2/\Gamma$ by putting
on it a D3-brane and its $|\Gamma|-1$ images ($|\Gamma| =$ the order
of $\Gamma$)\footnote{One needs $|\Gamma|$ images to make a full
  representation of $\Gamma$.} living in the 0123 dimensions.  Open
strings stretching between the branes provide a massless Neveu-Schwarz
sector that consists of the states
$$
\begin{array}{lll}
  A_\mu & \equiv \lambda_V \: \psi^\mu_{-1/2} |0\rangle_{NS} ,
  & \mu = 0,1,2,3 \\
  & \\
  x^i & \equiv \lambda_I \: \psi^i_{-1/2} |0\rangle_{NS} ,
  & i = 4,5 \\
  & \\
  x^m & \equiv \lambda_{II} \: \psi^m_{-1/2} |0\rangle_{NS} ,
  & m = 6,7,8,9
\end{array}
$$
where $\psi^\mu_{-1/2}$ are the lowest NS raising modes appearing in
the Laurent expansion of the string worldsheet fermions, and the
(tachyonic) NS ground state $|0\rangle_{NS}$ has odd fermion number,
$(-1)^F = -1$. The $|\Gamma| \times |\Gamma|$ Hermitian matrices
$\lambda_V$, $\lambda_I$ and $\lambda_{II}$ are Chan-Paton matrices,
required to obey invariance under $\Gamma$,
\begin{eqnarray}
\label{eq:lambdaV}
&  \gamma_\Gamma \lambda_V \gamma_\Gamma^{-1} = \lambda_V , & \\
\nonumber && \\
\label{eq:lambdaI}
&  \gamma_\Gamma \lambda_I \gamma_\Gamma^{-1} = \lambda_I , & \\
\nonumber && \\
\label{eq:lambdaII} &  \left(
  \begin{array}{c}
    \gamma_\Gamma \lambda_{II}^1\gamma_\Gamma^{-1} \\
    \\
    \gamma_\Gamma \lambda_{II}^2\gamma_\Gamma^{-1}
  \end{array}
  \right)
  = G_\Gamma \left(
  \begin{array}{c}
    \lambda_{II}^1 \\
    \\
    \lambda_{II}^2
  \end{array}
  \right) . &
\end{eqnarray}
Here the matrices $\gamma_\Gamma$ make up the regular representation
of the action of $\Gamma$ on the Chan-Paton indices, and $G_\Gamma$ is
some matrix in the $2\times 2$ representation of $\Gamma$, acting on
the two-vector $(\lambda_{II}^1,\lambda_{II}^2)$.

The invariance conditions (\ref{eq:lambdaV})--(\ref{eq:lambdaII})
break the original $U(|\Gamma|)$ gauge group to a product of unitary
subgroups, $F\equiv\prod_i U(N_i)$,\footnote{There is a diagonal
  $U(1)$ which acts trivially on vectors and hypermultiplets, implying
  that the actual nontrivial gauge group is $F/U(1)$.} and imply that
the gauge field on the brane, $A_\mu$, transforms in the adjoint of
$F$. Moreover, the $x^i$ make up a hypermultiplet in the adjoint of
$F$, and the $x^m$ make up a hypermultiplet transforming in the
fundamentals of subgroups $U(N_i) \times U(N_j)$, as (${\mathbf N}_i,
\overline{\mathbf N}_j$).\footnote{For this reason the $x^m$
  hypermultiplets are sometimes referred to as bifundamentals.}  The
$x^i$ may be interpreted as the position of a brane in the 45
directions, and the $x^m$ similarly parameterise motions of the brane
in the 6789 directions.

\begin{figure}[ht]
  \epsfxsize=9cm
  \centerline{\epsfbox{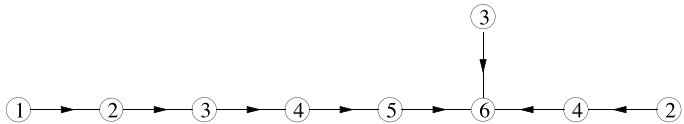}}
  \caption{The $E_8$ extended Dynkin diagram.}
  \label{fig:quiver}
\end{figure}

For $\Gamma = {\cal I}$, the gauge group is $F=U(1)\times U(2)^2\times
U(3)^2\times U(4)^2\times U(5)\times U(6)$, and the $x^m$
hypermultiplet\footnote{More correctly, in N=1 language the
  hypermultiplet is given by a pair $(\Phi,\overline{\Phi})$ of a
  chiral and an antichiral superfield. The chiral field transforms as
  described in the text and the antichiral transforms in the conjugate
  representation.} transforms as $({\mathbf 1}, \overline{\mathbf 2})
\oplus ({\mathbf 2}, \overline{\mathbf 3}) \oplus ({\mathbf 3},
\overline{\mathbf 4}) \oplus ({\mathbf 4}, \overline{\mathbf 5})
\oplus ({\mathbf 5}, \overline{\mathbf 6}) \oplus ({\mathbf 2},
\overline{\mathbf 4}) \oplus ({\mathbf 4}, \overline{\mathbf 6})
\oplus ({\mathbf 3}, \overline{\mathbf 6})$.  Comparing this
information with an $E_8$ Dynkin diagram, we see that the matter and
gauge content can be summarised by an extended Dynkin diagram, as in
\fig \ref{fig:quiver} \cite{DM}.  We associate each subgroup $U(N_i)$
with a node, letting the edges between them represent the $x^m$
hypermultiplets.  In addition, the edges are equipped with arrows to
indicate the way in which the hypermultiplets transform, the direction
being from the fundamental towards the antifundamental
representation.\footnote{Notice that a change of direction of an arrow
  only affects the final result by a change of sign in the
  corresponding FI parameter.}  The result is a \emph{quiver} diagram
\cite{DM}, which, besides being a convenient gauge theory summary in
general, will be useful to us when computing the curve that describes
the Higgs branch (in Section \ref{sec:E8curve}).

Due to the orbifolding of the 6789 directions one needs to take into
account twisted states.  We denote by a 3-vector $\vec{\phi}_k$ the
triplet of NS-NS twisted sector scalar fields associated with the
$k$:th $U(1)$ generator of $F$, \ie $k=1,...,9$ labels the nodes of
the quiver diagram. The low energy effective action on a D3-brane
then includes the potential term
\begin{eqnarray}
\nonumber V & \equiv & \sum_{i,j} \Tr ([x^i,x^j]^2)
+ \sum_{i,m} \Tr ([x^i,x^m]^2) \\
&& +\sum_{m,n} \Tr ([x^m,x^n]^2) + H(x^m,x^n,\vec{\phi}_k),
\label{eq:potential}
\end{eqnarray}
where $H$ is a function of $\vec{\phi}_k$ and the $x^m$
hypermultiplets, involving a term $(\vec{\phi}_k)^2$ and products of
the form $\phi_k^\alpha x^m x^n$ ($\alpha$ labels the three components
of $\vec{\phi}_k$). The last two terms of (\ref{eq:potential}) may be
rewritten using the definition \cite{JM}
\begin{equation}
  \vec{\mu}_k \equiv
  \Tr \left[ \lambda_V^k \left \{
      \varphi^{1\dag} \vec{\sigma} \varphi^{1} +\varphi^{2\dag}
      \vec{\sigma} \varphi^{2} \right \} \right] ,
\label{eq:momentmap}
\end{equation}
where $\varphi^{1}=(z^1,-\bar{z}^{\bar{2}})$ and
$\varphi^{2}=(z^2,\bar{z}^{\bar{1}})$, with $z^1 \equiv x^6+ix^7$ and
$z^2 \equiv x^8+ix^9$. The $\lambda_V^k$ are generators of the $k$:th
$U(1)$ subgroup of $F$, and $\vec{\sigma} \equiv
(\sigma^1,\sigma^2,\sigma^3)$ are the Pauli matrices.  Labelling the
rest of the generators of $F$ by the index $a$, we obtain
\begin{equation}
  \label{eq:HMpot}
  \sum_k (\vec{\mu}_k - \vec{\phi}_k)^2 + \sum_a (\vec{\mu}_a)^2
\end{equation}
for the last two terms of \eq (\ref{eq:potential}).

The vacuum condition $V=0$ yields the vacuum moduli space as two
branches, the Coulomb branch and the Higgs branch.  On the Coulomb
branch we have $\vec{\phi}_k=0$, $x^m=0$ and $x^i \neq 0$, \ie the
D3-brane is stuck at the orbifold singularity in the 6789 directions
but is free to move in the 45 directions. The Higgs branch, on the
other hand, requires $x^i=0$, implying that the D3-brane is stationary
in the 45 directions, while the $x^m$ may be nonzero. The kind of
space that the $x^m$ describe depends on the value of $\vec{\phi}_k$. If
$\vec{\phi}_k=0$ the Higgs branch is just the orbifold $\C^2/\Gamma$,
whereas for non-vanishing $\vec{\phi}_k$ this hypermultiplet space is a
resolved version of $\C^2/\Gamma$, with $\vec{\phi}_k$ controlling the
size of the singularity blow-up. The latter space is an Asymptotically
Locally Euclidean (ALE) space.  That the Higgs branch takes this form
may be seen by noting that the functions (\ref{eq:momentmap}) are
precisely the moment maps arising in the mathematical construction of
hyperk\"ahler quotients \cite{LR,HKLR87}.

It turns out that the twisted sector moduli $\vec{\phi}_k$ discussed
above are just the Fayet-Iliopoulos (FI) parameters of the quiver
theory.  The classical vacua in gauge theories are determined by
integrating out the auxiliary fields and requiring that the resulting
potential vanish. In our case one obtains, for each $U(1)$ generator
of the gauge theory, a real equation from the D-terms and a complex
equation from the F-terms. These together constitute the three
equations
\begin{equation}
  \label{eq:susycnd}
  \vec{\mu}_k - \vec{\zeta}_k = 0,
\end{equation}
where $\vec{\mu}_k$ is defined as in \eq (\ref{eq:momentmap}) and the
3-vector $\vec{\zeta}_k \equiv (\zeta^R,\zeta^C,\overline{\zeta^C})$
denotes the triplet of FI terms associated with the $k$:th $U(1)$
generator (R stands for ``real'' and C for ``complex''). The condition
(\ref{eq:susycnd}) is precisely the vacuum requirement that the first
term in \eq (\ref{eq:HMpot}) vanish, if we identify $\vec{\phi}_k$
with $\vec{\zeta}_k$, an identification which is corroborated in
Section \ref{sec:HWpicture}.

Thus we see that the FI parameters are associated with resolution of
the orbifold singularity, and that there is one FI term for every
node of the quiver diagram. In fact, the FI parameters may
be defined as the period of the hyperk\"ahler triplet $\vec{\omega}$
of complex forms,
\begin{equation}
  \label{eq:period}
  \vec{\zeta}_k  \equiv \int_{\Omega_k} \vec{\omega} ,
\end{equation}
where $\Omega_k$ is the $k$:th of the eight\footnote{The fact that one
  $U(1)$ acts trivially leads to a relation among the FI parameters,
  see Section \ref{sec:E8curve}.} 2-cycles required to blow up the
$\C^2/{\cal I}$ singularity.\footnote{The $\C^2/\Gamma$ singularities
  fall into an ADE classification according to the pattern of
  intersecting 2-cycles required to resolve them \cite{Aspinwall}.
  These 2-cycles behave exactly like simple roots of the corresponding
  Dynkin diagram, and the classification matches the quiver diagrams
  so that the 2-cycles may be identified with the FI parameters
  through \eq (\ref{eq:period}).}

\sect{The E$_8$ calculations}
\label{sec:E8curve}

The algebraic variety describing the Higgs branch of the quiver theory
is defined by an equation involving three polynomials in the
bifundamentals, that are invariant under the gauge group $F$
\cite{JM}.  Our goal here is to show that this curve is identical to
the Seiberg-Witten (SW) curve describing the generalised Coulomb
branch of the SW theory with $E_8$ global symmetry.  The
generalised Coulomb branch was defined in \cite{E7} as the fibration
of the Seiberg-Witten torus over the ordinary Coulomb branch. The SW
curve was found in \cite{MN2}, but we will use the results of
\cite{Noguchi}, as their form of the curve is more useful for our
purposes.

\begin{figure}[ht]
  \epsfxsize=13cm \centerline{\epsfbox{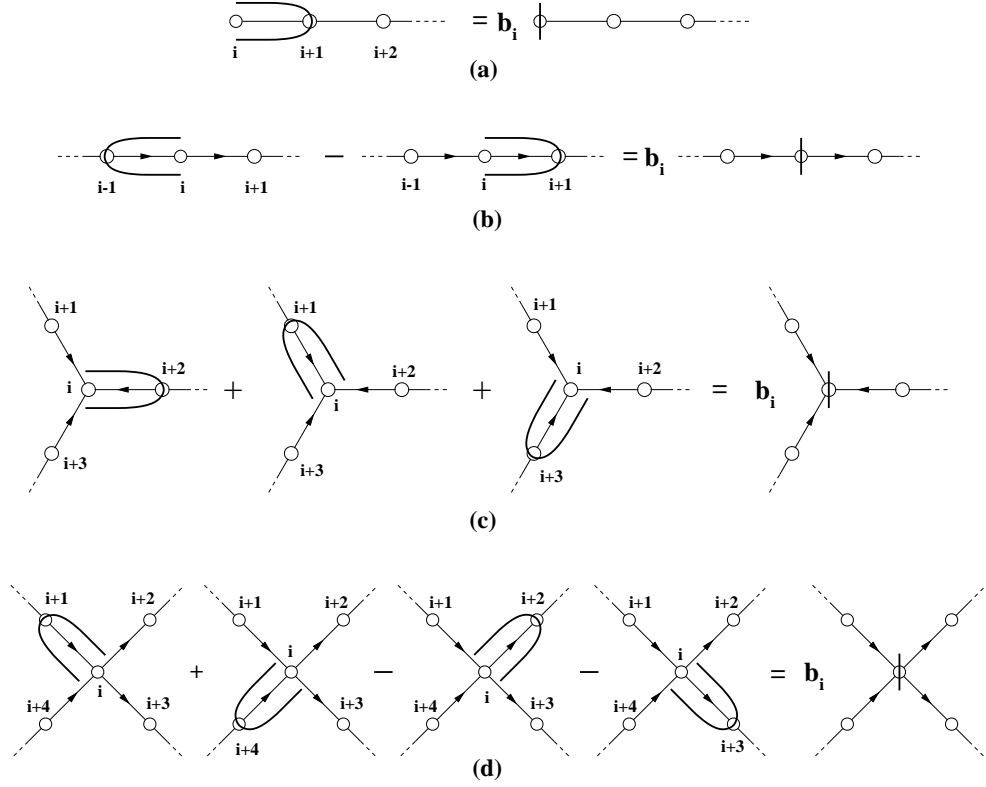}}
  \caption{
    The rules to manipulate traces in a quiver diagram. The $b_i$ are
    Fayet-Iliopoulos parameters.}
  \label{fig:rules}
\end{figure}

To find the algebraic curve we use the graphical method described in
\cite{E6}, which is based on the quiver diagram. Traces over the
bifundamentals are represented by loops in the diagram, and the
F-flatness conditions (\ref{eq:susycnd}) are imposed as graphical
rules for manipulating these traces.  The rules are summarised in \fig
\ref{fig:rules}.
Note that, by combining traces of these rules, we obtain a relation
between the FI terms,
\begin{equation}
b_1 - 2b_2 - 3b_3 - 4b_4 - 5b_5 - 6b_6 - 4b_7 + 2b_8 + 3b_9 = 0,
\label{eq:FIrel}
\end{equation}
which allows us to eliminate one of the $b_i$'s.

\begin{figure}[ht]
  \epsfxsize=9cm
  \centerline{\epsfbox{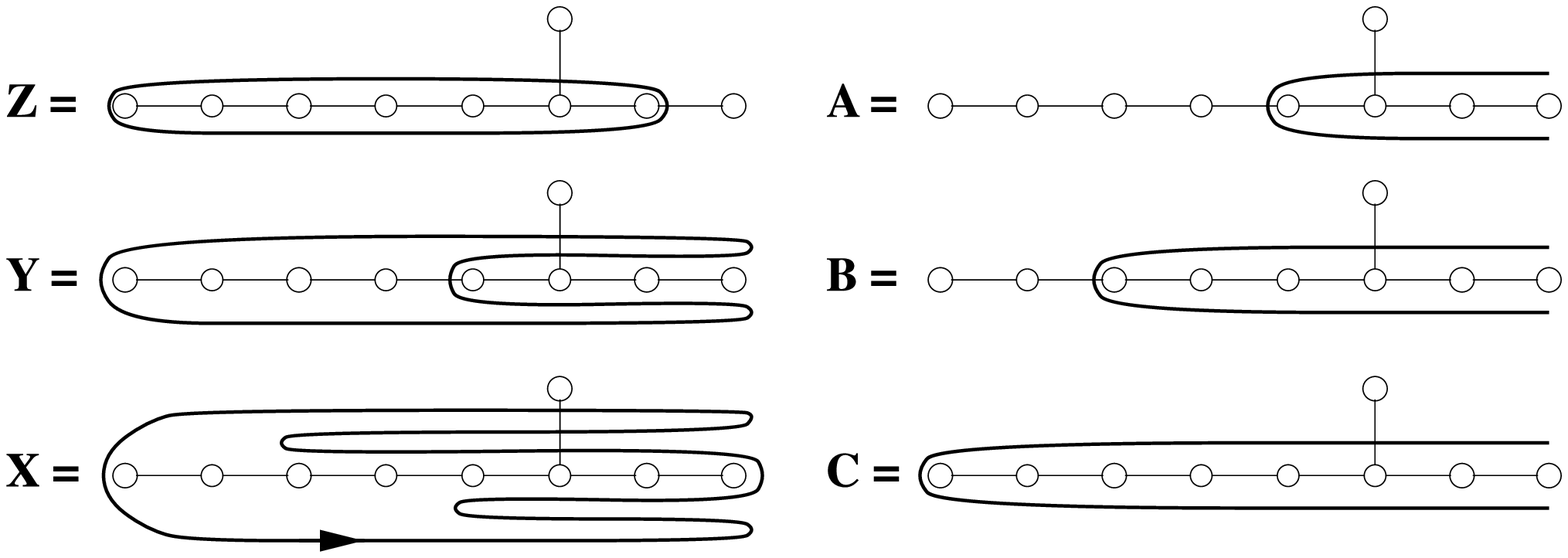}}
  \caption{Invariants and matrices.}
  \label{fig:e8defs}
\end{figure}

To ultimately obtain a form of the curve that allows immediate
comparison with the SW curve as given in \cite{Noguchi}, our approach
is to express the square of the highest-order invariant in terms of
lower-order invariants.  For this we need the three $F$-invariants
$X$, $Y$ and $Z$, as well as some simpler matrices $A$, $B$ and $C$
that we define to simplify calculations; they are all given in \fig
\ref{fig:e8defs}.

Using the Schouten identity
$$
\Tr(\{A,B\}C)=\Tr(AB)\Tr(C)+\Tr(AC)\Tr(B)
$$
$$
+\Tr(BC)\Tr(A)-\Tr(A)\Tr(B)\Tr(C)
$$
and the cyclic property of the trace we can reduce the square of
the highest invariant, which, as can be seen from the diagram (using a
$U(1)$ Schouten identity), may be written as $X^2=\Tr(ABCABC)$, to
products of ($X$ and) traces of at most two of the matrices $A$, $B$,
$C$,\footnote{Remark: For the Schouten identity to be useful for
  traces of an odd number of matrices, it is necessary that at least
  two of them are the same.}
\begin{eqnarray}
\nonumber &&X^2= \\ \nonumber
       && 2\Tr(ABC)\left( \myphantom{0.5cm} \Tr(A)\Tr(BC)+
          \Tr(B)\Tr(AC) \right. \\ \nonumber
       && \left. \myphantom{0.5cm} +\Tr(C)\Tr(AB)
          -\Tr(A)\Tr(B)\Tr(C) \right) \\ \nonumber
       && +\Tr(A^2)\left( \Tr(BC)^2+\Tr(B)\Tr(C)
          \left[\half \Tr(B)\Tr(C)-\Tr(BC) \right]\right) \\\nonumber 
       && +\Tr(B^2)\left( \Tr(AC)^2+\Tr(A)\Tr(C)
          \left[\half \Tr(A)\Tr(C)-\Tr(AC) \right]\right) \\\nonumber 
       && +\Tr(C^2)\left( \Tr(AB)^2+\Tr(A)\Tr(B)
          \left[\half \Tr(A)\Tr(B)-\Tr(AB) \right]\right) \\\nonumber 
       && -\half \Tr(A^2)\Tr(B^2)\Tr(C^2)+\\\nonumber
       && +\Tr(A)\Tr(B)\Tr(C)\left( \myphantom{0.5cm} 
          \Tr(A)\Tr(BC)+\Tr(B)\Tr(AC)+\Tr(C)\Tr(AB) \right) \\\nonumber
       && \myphantom{0.7cm} -\Tr(A)^2\Tr(BC)^2-\Tr(B)^2\Tr(AC)^2-\Tr(C)^2
       \Tr(AB)^2 \\
       && \myphantom{0.7cm} -\Tr(A)^2\Tr(B)^2\Tr(C)^2-2\Tr(AB)\Tr(AC)\Tr(BC)
       \label{eq:tracecurve}
\end{eqnarray}

Having done this, we realize that in fact we need only to calculate a
few traces. However, these calculations are nontrivial and the results
are much too lengthy to fit in this paper,\footnote{All results can be
  downloaded from the web, in Maple V format, at
  http://www.physto.se/\~{}brinne/E8quiver/.} so we only give the
general form of the traces, as polynomials in $X$, $Y$ and $Z$,
\begin{equation}
  \begin{array}{rcl}
    \Tr(A)   &=& k_1 \\
    \Tr(B)   &=& k_2 \\
    \Tr(A^2) &=& -2Z + k_3  \\
    \Tr(C)   &=& k_4 Z+ k_5 \\
    \Tr(AB)  &=& k_6 Z+ k_7  \\
    \Tr(AC)  &\equiv& Y  \\
    \Tr(B^2) &=& -2Y + k_8 Z + k_9  \\
    \Tr(BC)  &=& Z^2 + k_{10} Y + k_{11} Z + k_{12}  \\
    \Tr(C^2) &=& \Tr(C)^2  \\
    \Tr(ABC) &=& X
  \end{array}
\label{eq:traces}
\end{equation}
The coefficients $k_1$, $k_2$, etc., are polynomials in the
Fayet-Iliopoulos terms $b_i$, $i=2,...,9$.\footnote{$b_1$ was
  eliminated using \eq (\ref{eq:FIrel}).}

We note here that the invariants $X$, $Y$ and $Z$ are of order 15, 10
and 6 in the $b_i$'s, and the traces of $A$, $B$ and $C$ are of order
3, 5 and 7, respectively.  From this information it is easy to find
the orders of the coefficients defined in (\ref{eq:traces}); for
instance, the ones that are most difficult to compute,\footnote{When
  calculating $k_9$, we found it was necessary to use the Schouten
  identity to express $\Tr(A^3)=\half
  \Tr(\{A,A\}A)=\frac{3}{2}\Tr(A)\Tr(A^2)-\half \Tr(A)^2$.} $k_9$ and
$k_{12}$, are polynomials of order 10 and 12, respectively, in eight
variables.

To compare our curve (\ref{eq:tracecurve}) with the SW curve, we need
to put it on its canonical form,
\begin{equation}
  X^2  = -Y^3+f(Z)Y+g(Z) , 
\label{eq:E8curve}
\end{equation}
where
\begin{eqnarray*}
  f(Z) &=& \omega_2 Z^3 + \omega_8 Z^2 + \omega_{14} Z + \omega_{20} , \\
  g(Z) &=& -Z^5 + \omega_{12} Z^3 + \omega_{18} Z^2 +
                  \omega_{24} Z + \omega_{30} ,
\end{eqnarray*}
and the $\omega_n$ are polynomials of order $n$ in the $b_i$'s. We
accomplish this by plugging in the traces (\ref{eq:traces}) into \eq
(\ref{eq:tracecurve}) and shifting our variables $X$, $Y$, $Z$.
The substitution of traces puts our curve on the form
\begin{eqnarray*}
X^2 - K_X X &=& -Y^3 -Z^5 + (\alpha_{Y^2Z}Z+\alpha_{Y^2})Y^2 \\
&& +(\alpha_{YZ^3}Z^3+\alpha_{YZ^2}Z^2+\alpha_{YZ}Z+\alpha_{Y})Y \\
&& +\alpha_{Z^4}Z^4+\alpha_{Z^3}Z^3+\alpha_{Z^2}Z^2+\alpha_{Z}Z+\alpha_0,
\end{eqnarray*}
where
\begin{eqnarray*}
K_X & \equiv & \beta_{Y}Y+\beta_{Z^2}Z^2+\beta_{Z}Z+\beta_0
\end{eqnarray*}
and the $\alpha$'s and $\beta$'s are independent of the invariants
$X$, $Y$ and $Z$.  To get rid of the $X$-term, we shift $X$ by $-\half
K_X$. Expanding the result, we obtain a coefficient of the $Y^2$-term
such that we must shift $Y$ by $-\third (\quart
\beta_Y^2+\alpha_{Y^2Z}Z+\alpha_{Y^2})$ in order to eliminate $Y^2$.
Finally, the $Z^4$-term is substituted away by $Z \rightarrow Z-\fifth
(\quart \beta_{Z^2}^2+\third
\alpha_{YZ^3}\alpha_{Y^2Z}+\alpha_{Z^4})$, and we end up with unwieldy
expressions for the $\omega_n$'s, which should be compared to the
Seiberg-Witten coefficients $w_n$ of \cite{Noguchi}.

Finding the explicit relation between our FI parameters and the mass
parameters $w_n$ would be a next to impossible task unless we had a
good idea of what it should be. Guided by our conjecture that the
Higgs branch algebraic curve and the Seiberg-Witten curve are
identical, and the fact that Noguchi et~al.~\cite{Noguchi} used
Casimir invariants to express their SW curve, we use $E_8$ Casimirs
written in terms of the FI parameters.  It is then straightforward,
although very time consuming, to compare our coefficients with those
of \cite{Noguchi}.

To find the Casimirs, first note that the simple roots of $E_8$ may,
as we saw in Section \ref{sec:Quiver}, be identified with the FI
parameters via \eq (\ref{eq:period}),
$$
  b_i = \int_{\Omega_i} J ,
$$
where $\Omega_i$ is a 2-cycle behaving as a simple root, and $J$ is
the K\"ahler form on the orbifold. The Casimir invariants $P_k$ may be
found as \cite{Noguchi}
$$
P_k = 2^{-k} c_{240-k} ,
$$
where $c_{240-k}$ is the coefficient of $t^{248-k}$ in the
characteristic polynomial
\begin{equation}
  \det (t-v\cdot H) = \prod_{k=1}^{248} (t-v_k) ,
\label{eq:charpoly}
\end{equation}
where $v_k$ are the weights of the fundamental representation, and
$v\cdot H \equiv \diag (v_1, ..., v_{248})$.  Factoring out $t^{248}$
and defining
\begin{equation}
  \chi_n \equiv \Tr \left[ (v\cdot H)^n \right]
  = \sum_{k=1}^{248} v_k^n ,
  \label{eq:chi}
\end{equation}
we may rewrite (\ref{eq:charpoly}) as
$$
\begin{array}{c}
t^{248} \prod_{k=1}^{248} (1- {v_k \over t}) 
= t^{248} \exp \left\{ \sum_{k=1}^{248} \ln \left( 1- {v_k \over t}
\right) \right\} \\
\\
= t^{248} \sum_{m=0}^{\infty} \left( -{1\over m!} \right)^m
\left( \sum_{n=1}^{\infty} {\chi_n \over n t^n} \right)^m ,
\end{array}
$$
a form which facilitates the extraction of coefficients.  Writing
the weights $v_k$ in the usual way as linear combinations of simple
roots (see Appendix \ref{sec:weights}), we thus obtain the Casimirs
$P_k$ expressed in FI parameters $b_i$.\footnote{The trace identities
  listed in Appendix \ref{sec:traceids} were useful in this
  calculation.}

To actually compare $\omega_n$ and $w_n$ turned out to be quite
demanding. The highest-order coefficient $\omega_{30}$ is a polynomial
of order 30 in eight variables, which means that it has in principle
10295472 terms. We did our calculations in Maple V software on a
Digital Alpha workstation, but since it would not store a polynomial
of that size in its memory, much less compare two of them, we tried to
find other software that would, without success.  Below we describe
the partly numerical check we performed, which, short of writing a
dedicated program, seems the best one possible, as well as totally
convincing.

We explicitly checked up to $\omega_{12}$ by comparing the full
expressions for the polynomials and got complete agreement. The higher
order $\omega$'s we could only check numerically and they agree when
all the $b_i$'s are set to random and different prime numbers.  We
have thus shown that the $E_8$ quiver Higgs branch is equal to the
generalised Coulomb branch of the SW theory with $E_8$ global
symmetry.

\section{The Hanany-Witten picture}
\label{sec:HWpicture}

The IIB picture of D3-branes on a $\C^2/\Z_n$ orbifold singularity
(which is of type $A_{n-1}$) is T-dual to a picture of type IIA string
theory in a background of D4-branes stretching between NS5-branes
\cite{KLS}. This dual picture, the Hanany-Witten (HW) picture
\cite{HW}, provides an intuitive geometric interpretation of blow-ups
of $A_{n-1}$ type singularities. An analogous picture exists for $D_n$
type singularities \cite{Kapustin,HZ}, and it seems plausible that
there are generalisations also to $E_6$, $E_7$ and $E_8$.  In this
section, we analyse the HW picture for the $\C^2/\Z_n$ case along the
lines of \cite{KLS} (see also \cite{Dasgupta2,Johnson}); in particular
we clarify the role of the Fayet-Iliopoulos terms.

Starting from the type IIB string theory configuration ($\times$ means
the object is extended in that direction, and $-$ means it is
pointlike)
\begin{eqnarray*}
&&  \begin{array}{|c|c|c|c|c|c|c|c|c|c|c|}
         \hline
         & $0$ & $1$ & $2$ & $3$ & $4$ & $5$ & $6$ & $7$ & $8$ & $9$ \\\hline
    \mathrm{sing} & \times & \times & \times & \times & \times & \times & - & - & - & - \\\hline
    \mathrm{D}3   & \times & \times & \times & \times & - & - & - & - & - & - \\\hline
  \end{array}
\end{eqnarray*}
we T-dualise along the 6-direction to get
\begin{eqnarray*}
&&  \begin{array}{|c|c|c|c|c|c|c|c|c|c|c|}
         \hline
         & $0$ & $1$ & $2$ & $3$ & $4$ & $5$ & $6$ & $7$ & $8$ & $9$ \\\hline
    \mathrm{NS}5  & \times & \times & \times & \times & \times & \times & - & - & - & - \\\hline
    \mathrm{D}4   & \times & \times & \times & \times & - & - & \times & - & - & - \\\hline
  \end{array}
\end{eqnarray*}
in type IIA string theory in flat spacetime. There are $n$ NS5-branes,
which all coincide in the 789 directions, but not necessarily in the
6-direction.  Between them D4-branes are suspended, which are the
T-duals of the IIB D3-branes.  The rotational symmetry $SO(3) \simeq
SU(2)$ of the 789 coordinates translates into the $SU(2)_R$ symmetry
of the gauge theory living on the D4-branes.  The hypermultiplets
arise from fundamental strings stretching across the NS5-branes,
between neighbouring D4-branes.

Resolving singularities in the IIB picture corresponds to separating
NS5-branes along the 789 directions in the IIA picture. By an $SU(2)$
rotation we can always pick the direction of displacement to be $x^7$.
Note that such a displacement breaks the 789 rotational symmetry; that
is, blowing up a singularity breaks the $SU(2)_R$ symmetry.
If we move some of the NS5-branes in this way, with
the D4-branes still stuck to them, and then T-dualise along $x^6$
again, we do not regain the D3-brane picture. Rather, the now tilted
D4-branes dualise to a set of D5-branes (with nonzero B-field) with
their 67 worldvolume coordinates wrapped on 2-cycles. Shrinking these
2-cycles to zero size, each of the wrapped D5-branes is a fractional
D3-brane, which cannot move away from the singularity.  Thus a
fractional D3-brane corresponds to a D4-brane whose ends are stuck on
NS5-branes.

To move a fractional D3-brane, or, equivalently, a wrapped D5-brane,
along the 6789 directions, we need to add $n-1$ images (under $\Z_n$),
all associated with a 2-cycle each. The sum of the full set of
2-cycles is homologically trivial and can be shrunk to zero size. Then
the collection of wrapped D5-branes will look like a single D3-brane
that can move around freely in the orbifold.  This procedure
corresponds in the HW picture to starting out with a single D4-brane
stretching between two of the $n$ NS5-branes, and wanting to move the
D4-brane (in the 7-direction, say) away from the NS5-branes, detaching
its ends. In order not to violate the boundary conditions of the
D4-brane, we then need to put one D4-brane between each unconnected
pair of NS5-branes and join them at the ends. We then get a total of
$n$ D4-branes forming a single brane winding once around the periodic
6-direction.  The D4-brane may now be lifted off the NS5-branes and
move freely, corresponding to the free D3-brane in the T-dual picture.

We may also gain some insight concerning the role played by the FI
parameters in the HW picture, from the worldvolume theory of a wrapped
D5-brane on the orbifold singularity. Consider such a brane living in
the 012367 directions, with its 67 worldvolume coordinates wrapped on
a 2-cycle $\Omega_k$.  The Born-Infeld and Chern-Simons terms in the
worldvolume action are, schematically, \cite{DM}
\begin{eqnarray}
  \nonumber I_{D5} &=& \int d^6x \sqrt{\det(g+{\cal F})}
  +\mu\int C^{(6)}
  +\mu\int C^{(4)}\wedge {\cal F} \\
  && +\mu\int C^{(2)}\wedge {\cal F}\wedge {\cal F}
  +\mu\int C^{(0)}\wedge {\cal F}\wedge {\cal F}\wedge {\cal F} ,
  \label{eq:D5action}
\end{eqnarray}
where $g$ is the metric on the worldvolume, $C^{(p)}$ is the R-R
$p$-form, $\mu$ is a constant, and ${\cal F}=F^{(2)}+B^{NS}$ where the
2-form $F^{(2)}$ is the field strength of the gauge field on the brane
and $B^{NS}$ is the NS-NS 2-form on the brane.  Dimensional reduction
to the 0123 directions, by integrating over the 2-cycle, puts the
first term of (\ref{eq:D5action}) on the form
\begin{equation}
  \int_{\Omega_k} d^2x \sqrt{\det( g_2+ {\cal F}_2)}
  \int d^4x \sqrt{\det( g_4+ {\cal F}_4)} ,
  \label{eq:D5BIred}
\end{equation}
where ${\cal F}_2 = C^{(2)}+B^{NS}$, $g_2$ is the metric on the 67
directions, and $g_4$ is the metric on the 0123 directions.  Expanding
(\ref{eq:D5BIred}) we obtain the coupling constant $g_k^{-2}$ in four
dimensions as the coefficient of $\int d^4x F_{\mu\nu}F^{\mu\nu}$.  It
is just the factor on the left in (\ref{eq:D5BIred}), which we can
write as
\begin{equation}
  g_k^{-2} = \left| \int_{\Omega_k} \left( B^{NS} + iJ \right) \right| .
  \label{eq:gk}
\end{equation}
In the HW picture the coupling constant of the four dimensional theory
is proportional to the length of the D4-brane in the additional fifth
direction of the brane. Hence (\ref{eq:gk}) measures the total
distance between two NS5-branes between which the D4-brane is
suspended. Furthermore, since the distance between the NS5-branes in
the isometry direction (in our case $x^{6}$) is given by the flux of
the $B^{NS}$ field on the corresponding cycle, we have to interpret
$\int_{\Omega_k} J$ as the position of the NS5-branes in a direction
orthogonal to that, let us choose $x^{7}$. Movement of the NS5-branes
in the remaining directions $x^{8}$ and $x^{9}$ now corresponds to
turning on the $SU(2)_R$ partners of the K\"{a}hler form.

The integral of $J$ over a 2-cycle is also, by definition, a
Fayet-Iliopoulos term.  A hyperk\"ahler manifold has an SU(2) manifold
of possible complex structures.  Choosing a complex structure we can
define the K\"{a}hler form $J$ as $\omega^1$, and the holomorphic
2-form as $\omega^2 +i\omega^3$. These three 2-forms rotate into each
other under $SU(2)_R$ transformations, corresponding to choosing a
different complex structure. The $k$:th triplet of FI terms is defined
by the period of $\vec{\omega} = (\omega^1,\omega^2,\omega^3)$ (and
hence also transforms as a triplet under $SU(2)_R$), as
$$
\vec{\zeta}_k  \equiv \int_{\Omega_k} \vec{\omega} .
$$
Hence
$$
\zeta^R_k = \int_{\Omega_k} J ,
$$
where $\zeta^R_k$ is the real component of the triplet of FI terms
$\vec{\zeta}_k = (\zeta_k^R,\zeta_k^C,\overline{\zeta_k^C})$.

Another way to obtain the FI terms of the four-dimensional Yang-Mills
theory is via dimensional reduction and supersymmetrisation of the
D5-brane worldvolume theory \cite{DM}.  The third term of
(\ref{eq:D5action}) can be rewritten as
$$
\int d^6x (A_\mu - \partial_\mu c^{(0)})^2,
$$
where $c^{(0)}$ is the Hodge dual potential of $C^{(4)}$ in six
dimensions.  After integration over the $k$:th 2-cycle we
supersymmetrise this to
$$
\int d^4x d^4\theta ({\mathbf C}_k - \overline{\mathbf C}_k -
{\mathbf V})^2,
$$
where ${\mathbf C}_k$ is a chiral superfield whose complex scalar
component is $c^{(0)} + i\zeta^R_k$, and $\mathbf V$ is the vector
superfield containing $A_\mu$.  Here the imaginary part $\zeta^R_k$ of
the scalar component is the real FI term in four dimensions, and
we see that it arises as the superpartner of $c^{(0)}$.

\section{Conclusions}
\label{sec:Conclusions}

In this paper we have discussed the conjectured mirror symmetry
between the Higgs branch of quiver gauge theories and the generalised
Coulomb branch of certain four-dimensional strongly coupled, globally
symmetric gauge theories. This symmetry was hinted at in \cite{E6},
where equivalence of the $E_6$ curves was established, and it was
substantiated in \cite{E7}, where equivalence of the $E_7$ curves was
presented and the conjecture was made precise. Here we have shown the
final equivalence, that of the $E_8$ curves for the two branches.

The proof we have given is by direct comparison.  It would be nice to
find a proof such as a chain of dualities leading from one model to
the other, as indicated in \cite{E7}.  There are obstacles to this in
that, e.g., many of the dualities expected to enter that chain are not
known.

In this context we emphasize an important aspect of our results,
namely as a guide for finding Hanany-Witten type constructions. Assume
that one is contemplating a HW picture of NS5-branes with D4-branes
ending on them and that this is supposed to describe the IIA T-dual of
D3-branes on an ($E_6$, $E_7$ or) $E_8$ singularity in IIB. Moving the
NS5-branes in the HW picture corresponds to blowing up the singularity
in the dual picture. As discussed in Section \ref{sec:HWpicture}, the
FI parameters give the position of the NS5-branes in the HW picture
when they are moved.  Since we have given the relation of the FI
parameters to the parameters governing the deformation of the
algebraic variety, one may now check that the possible motions on the
HW side (allowed by the particular geometry suggested) correspond to
the known allowed deformations.

We have performed the above check for some of the known dualities and
hope to use it in future efforts to find HW pictures of the
$E_n$-theories.

Finally we mention that all is set up for finding the remaining trace
identities in Appendix \ref{sec:traceids}. The calculations will be
carried out as soon as the computer capacity becomes available to us.

\section{Acknowledgements}

The work of UL was supported in part by NFR grant 650-1998368 and by
EU contract HPRN-CT-2000-0122. RvU was supported by the Czech Ministry
of Education under Contract No 143100006. He would also like to thank
Stockholm University for hospitality during part of this work.

\renewcommand{\thesection}{A}

\section{Appendix}

\subsection{The E$_8$ weights}
\label{sec:weights}

The $E_8$ Casimir invariants may be expressed in terms of fundamental
weights as shown in Section \ref{sec:E8curve}. The weights may in turn
be expressed in terms of the $E_8$ simple roots.\footnote{Although we
  treat the case $E_8$ here, the procedure is quite general.}

We want to find the 248 weights of the fundamental representation of
$E_8$. This is the same as the adjoint representation, and it is real,
whence half of the weights are just minus the first half. In addition
one finds that eight of them are zero, so we end up with only 120
nonvanishing independent weights.

Viewing the simple roots $b_i$ as 8-vectors, the Cartan matrix for
$E_8$,
\begin{eqnarray*}
&&  \left(
  \begin{array}{cccccccc}
     2 & -1 &  0 &  0 &  0 &  0 &  0 &  0 \\
    -1 &  2 & -1 &  0 &  0 &  0 &  0 &  0 \\
     0 & -1 &  2 & -1 &  0 &  0 &  0 &  0 \\
     0 &  0 & -1 &  2 & -1 &  0 &  0 &  0 \\
     0 &  0 &  0 & -1 &  2 & -1 &  0 & -1 \\
     0 &  0 &  0 &  0 & -1 &  2 & -1 &  0 \\
     0 &  0 &  0 &  0 &  0 & -1 &  2 &  0 \\
     0 &  0 &  0 &  0 & -1 &  0 &  0 &  2
  \end{array}
  \right) ,
\end{eqnarray*}
may be viewed as the matrix of scalar products $b_i \cdot b_j$. If we
start with a highest weight vector $w_0$, lower weights are obtained
by subtracting simple roots whose scalar products with $w_0$ are
positive \cite{Georgi}. To be explicit, choose $w_0$ such that its
scalar product with $b_1$ is 1 and otherwise zero, \ie $w_0$ would be
represented by a row $[1,0,0,0,0,0,0,0]$ in the Cartan matrix. Then
the only possible next weight is $w_0-b_1$, represented by
$[-1,1,0,0,0,0,0,0]$ in the Cartan matrix. Iterating this procedure,
subtracting simple roots whose scalar products with the previous
weight are positive, we obtain 120 nonvanishing weights $w_k$ in
``Cartan representation''. To express these in terms of simple roots,
we invert the Cartan matrix and compute $v_k$ as
$$
v_k = 2 \sum_{j=1}^8 \sum_{i=1}^8 w_k^j C_{ji} b_i
$$
where $w_k^j$ denotes the $j$:th component of the vector $w_k$, $C$
is the inverse of the Cartan matrix, and the factor 2 is a
normalisation.

Note the slight change in convention; all the $b_i$'s here have the
same sign, whereas two of our FI terms in Section \ref{sec:E8curve}
($b_8$ and $b_9$) have opposite sign relative to the others.

\subsection{Trace identities}
\label{sec:traceids}

The traces $\chi_n \equiv \Tr [(v\cdot H)^n]$ (\eq (\ref{eq:chi}))
used in writing the Casimir invariants in Section \ref{sec:E8curve}
satisfy identities that simplify the calculations slightly.  Such
trace identities were derived by \cite{Traceids} for the simple Lie
algebras up to $E_6$, whereas the $E_7$ identities were calculated by
the authors of \cite{E7}. For $E_8$ we found the following identities,
$$
\begin{array}{lll}
  \chi_4 &=& {1 \over 100} \chi_2^2 \\
  \chi_6 &=& {1 \over 7200} \chi_2^3 \\
  \chi_{10} &=& {1 \over 16} \chi_2 \chi_8 - {1 \over 69120000} \chi_2^5
\end{array}
$$
Although straightforward in principle, due to limitations of Maple,
we were not able to derive the rest of the trace identities, \ie those
expressing $\chi_{16}$, $\chi_{22}$, $\chi_{26}$ and $\chi_{28}$ in
terms of the independent $\chi_n$'s.

\end{document}